\documentclass[fp,twocolumn]{jpsj3}
\usepackage{txfonts}
\usepackage{color}

\title{Superconductivity of Electron-Doped NdOBiS$_{2}$ by Substitution of Mixed-Valence Ce ions}

\author{Naoki Kase$^1$\thanks{Author to whom correspondence should be addressed: n-kase@rs.tus.ac.jp}, Masaya Matsumoto$^1$, Katsuo Kondo$^1$, Jun Gouchi$^2$, Yoshiya Uwatoko$^2$, Toshiro Sakakibara$^2$, and Nobuaki Miyakawa$^1$}

\inst{
$^1$Department of Applied Physics, Tokyo University of Science, Tokyo 125-8585, Japan\\
$^2$Institute for Solid State Physics, University of Tokyo, Kashiwa 277-8581, Japan
} 

\abst{
Superconductivity is achieved in Nd$_{1-x}$Ce$_{x}$OBiS$_{2}$ via electron doping using mixed-valence Ce ions.
Single crystals with \textit{x} = 0.2, 0.3, 0.4, and 0.5 are generated using a CsCl flux method.
Plate-like single crystals with dimensions of 0.8$\times$0.8$\times$0.2 mm$^3$ were obtained.
The magnetic susceptibility $\chi(T)$ indicates large diamagnetism, and the electrical resistivity $\rho(T)$ indicates zero resistivity.
The maximum value of $T_\mathrm{c}$ is observed at 4.7 K in Nd$_{0.8}$Ce$_{0.2}$OBiS$_{2}$ from $\chi(T)$.
From the $\rho(T)$ measurements taken in several magnetic fields, the upper critical field $\mu_0H_\mathrm{c2}$(0) is estimated to be $\sim$12 and $\sim$0.34 T for the $ab$- and $c$-planes, respectively.
We redetermined $\mu_0H_\mathrm{c2}(0)$ of NdO$_{0.7}$F$_{0.3}$BiS$_2$, as $\sim$35 and $\sim$0.78 T for the $ab$- and $c$-planes, respectively.
The anisotropic parameter $\Gamma$ is estimated to be $\sim$35 for Nd$_{0.7}$Ce$_{0.3}$OBiS$_{2}$ and $\sim$45 for NdO$_{0.7}$F$_{0.3}$BiS$_2$.
The $\mu_0H_\mathrm{c2}$(0) of Nd$_{0.7}$Ce$_{0.3}$OBiS$_{2}$ is approximately two times smaller than that of NdO$_{0.7}$F$_{0.3}$BiS$_2$, although the difference of $\Gamma$ is approximately 10.
}

\begin{document}
\maketitle
The newly discovered BiS$_{2}$-based superconductors Bi$_{4}$O$_{4}$S$_{3}$ and RO$_{1-x}$F$_{x}$BiS$_{2}$ (R = La, Ce, Pr, Nd) are of interest owing to their various crystal structures similar to those in layered cuprate and iron pnictide superconductors\cite{1,2,3,4,5,6,7,8,9}.
The crystal structure of the BiCh$_{2}$-based (Ch: chalcogen) superconductors consists of alternating stacks of RO-blocking and BiCh$_{2}$ conducting layers.
It is known that electron carrier doping into the conduction band, mainly comprising Bi-6$p$ and Ch-$p$ orbitals\cite{theory1,theory2}, induces superconductivity in the BiS$_2$-based compounds.
Recent research has indicated that in-plane chemical pressure is related to the overlapping of the Bi-6$p$ and Ch-$p$ orbitals, which is a significant contributor to bulk superconductivity\cite{a0}.
A typical approach to increasing overlapping is by substituting elements with a longer ionic radius (e.g. Se substitution for S sites)\cite{a1,a2,a3}.
From detailed analysis of the crystal structure, the chemical pressure effects effectively ameliorated the in-plane disorder in the BiCh$_2$ conducting layer\cite{b0,b1,b2,b3}.

The mechanism of superconductivity in BiS$_{2}$-based compounds has received attention but remains under debate. 
The specific heat measurements of a bulk superconductor of LaO$_{0.5}$F$_{0.5}$BiSSe suggest full gap symmetry and indicate that superconductivity is mediated by electron--phonon coupling\cite{spe,spe1}.
However, the Se isotope effect suggests that the superconducting pairing interaction may not be due to electron--phonon interaction\cite{iso}.
In a single crystal of Nd(O,F)BiS$_{2}$, penetration depth and thermal conductivity indicate that superconductivity is a full-gap symmetry\cite{thermal1,thermal2,thermal3}.
In contrast, recent angle-resolved photoemission spectroscopy (ARPES)  measurements suggest that the superconducting gap of NdO$_{0.71}$F$_{0.29}$BiS$_{2}$ is anisotropic\cite{ARPES}.
Thus, the mechanism of the BiS$_{2}$-based superconductor is still controversial.

As previously mentioned, electron doping is necessary to induce bulk superconductivity in BiS$_{2}$-based compounds.
Recently, superconductivity has been observed in non-substituted CeOBiS$_2$; this is mostly because of electron doping by mixed-valence Ce ions\cite{CeOBiS2}.
The compound is considered to have the mixed valence of Ce$^{3+}$/Ce$^{4+}$.
In addition, superconductivity in the non-F-substituted LaOBiSSe and PrOBiS$_2$ is achieved by Ce-substitution at R-sites, suggesting that the carriers are injected owing to the mixed valence of Ce$^{3+}$/Ce$^{4+}$\cite{LaCe,PrCe}.
These results indicate that substitution by mixed-valence Ce ions can effectively induce superconductivity.

\begin{figure}
\begin{center}
\includegraphics[width=3.3in]{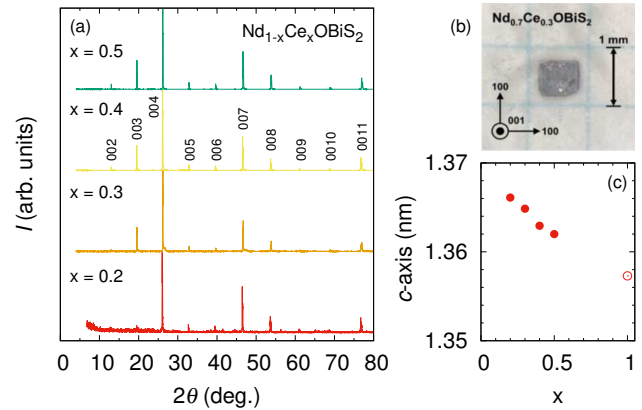}
\caption{(Color Online)
(a) X-ray diffraction patterns of Nd$_{1-x}$Ce$_{x}$OBiS$_2$ single crystals aligned along the (00$l$) plane with Cu-K$\alpha$ radiation.
The Miller indices of each peak are represented in the figure.
(b) A photograph of a piece of grown single crystal.
One grid spacing in the photograph represents 1.0 mm.
(c) Lattice constant of the $c$-axis of Nd$_{1-x}$Ce$_{x}$OBiS$_2$ at room temperature .
The data for CeOBiS$_2$ is extracted from Ref. 27.
}
\label{fig1}
\end{center}
\end{figure}

The investigation of superconductivity induced by electron doping except for F helps in a better understanding of the mechanism. 
Because F-substituted NdOBiS$_2$ is known to exhibit the highest superconducting transition temperature ($T_\mathrm{c}$) in the ROBiS$_2$ compounds at ambient pressure\cite{5,NdOBiS2}, a higher $T_\mathrm{c}$ is highly desirable in NdOBiS$_2$ by Ce-substitution.
In the present study,  we have successfully generated single crystals of Nd$_{1-x}$Ce$_{x}$OBiS$_{2}$ and reported the discovery of superconductivity in Nd$_{1-x}$Ce$_{x}$OBiS$_{2}$ through electrical resistivity $\rho(T)$ and magnetic susceptibility $\chi(T)$ measurements.
From the results, we revealed the superconducting properties of Nd$_{1-x}$Ce$_{x}$OBiS$_{2}$.
We noted that $T_\mathrm{c}$ of the crystals grown using the 99.9\%- and 99.999\%-CsCl fluxes were similar\cite{supp}.
In other words, we confirmed that the F impurity in 99.9\%-CsCl flux is not the primary factor in electron doping to induce superconductivity.

\begin{figure}
\begin{center}
\includegraphics[width=3.3in]{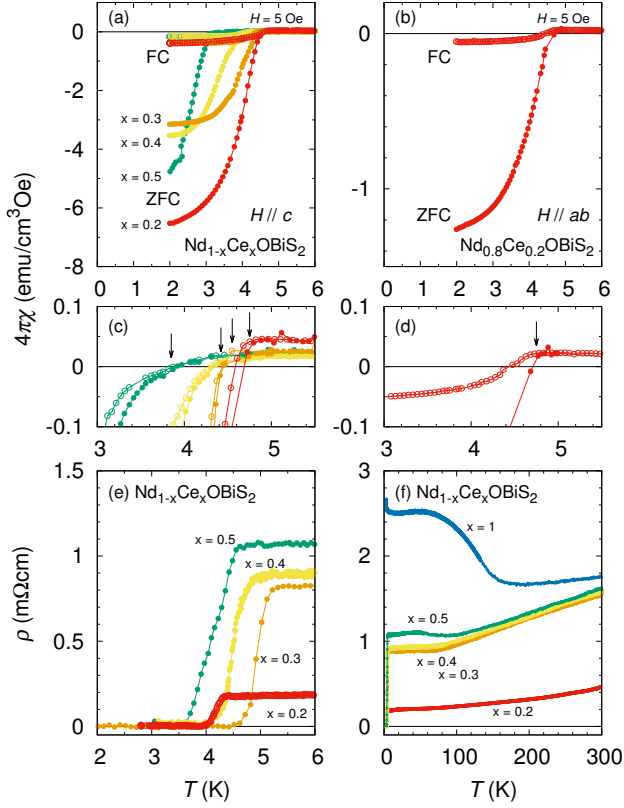}
\caption{(Color Online)
$T$-dependence of the magnetic susceptibility $\chi(T)$ of Nd$_{1-x}$Ce$_{x}$OBiS$_2$.
Zero-field-cooled (ZFC) and field-cooled (FC) measurements were performed with $H$ = 5 Oe applied along the (a) $c$- and (b) $ab$-plane.
(c-d) Enlarged parts of $\chi(T)$ around the superconducting transition.
(e-f) $T$-dependence of the electrical resistivity of Nd$_{1-x}$Ce$_{x}$OBiS$_2$.

}
\label{fig2}
\end{center}
\end{figure}

Single crystals of Nd$_{1-x}$Ce$_{x}$OBiS$_2$ with \textit{x} = 0.2, 0.3, 0.4, and 0.5 were synthesized by a CsCl-flux method.
The nominal compositions of Nd$_{2}$O$_{3}$ (99.9\%, Kojundo Chemical Lab. Co.), Nd$_{2}$S$_{3}$ (99\%, Kojundo Chemical Lab. Co.), CeO$_{2}$ (99\%, Kojundo Chemical Lab. Co.), and Bi$_{2}$S$_{3}$ (99.9\%, Kojundo Chemical Lab. Co.) were weighed (total mass of 0.8 g) and mixed with 5 g of CsCl powder (99.999\%,  Kojundo Chemical Lab. Co.).
The mixture was sealed in an evacuated quartz tube.
The quartz tube underwent heating at 950 ℃ for 10 h followed by cooling to 650 ℃ at a rate of 1 ℃/h; then, the sample was cooled to room temperature  in the furnace.
We successfully synthesized single crystals of Nd$_{1 - x}$Ce$_{x}$OBiS$_2$ with \textit{x}  = 0.2, 0.3, 0.4, and 0.5.
As shown in Fig. \ref{fig1}(b), the plate-like crystals were grown with \textit{x} = 0.3, 0.4, and 0.5, and the typical dimensions are approximately 0.8$\times$0.8$\times$0.2 mm$^{3}$.
The shape of the generated crystal with \textit{x} = 0.2 is slightly lumpy and small.
A single crystal with \textit{x} = 0.1 was not generated at this stage.
The results indicate that single crystals with low carrier doping are difficult to generate, and non-doped NdOBiS$_2$ could not be successfully synthesized\cite{5,NdOBiS2}.

The crystal structure was verified through X-ray diffraction (XRD), which employed a conventional X-ray spectrometer equipped with Cu-K$\alpha$ radiation (RINT 2500, Rigaku). 
The chemical composition of the single crystals of Nd$_{1-x}$Ce$_{x}$OBiS$_2$ was analyzed by X-ray fluorescence (XRF), conducted using a JEOL JSX 1000S ElementEye.
Electrical resistivity $\rho(T)$ was measured via a standard dc-four-probe method.
Electrical connections with the sample comprised gold wires ($\phi$ = 25 $\mu$m) joined with Ag paste (4922N, DuPont).
DC-magnetic susceptibility $\chi(T)$ was measured by a magnetic property measurement system (MPMS, Quantum Design), down to 2.0 K.
AC-electrical resistivity $\rho(T)$ was measured by a physical property measurement system (PPMS, Quantum Design), down to 2.0 K in magnetic fields of several intensities, including zero.
Thermoelectric power $S$ was measured using a seesaw heating method by considering a different temperature $\Delta T$ of about 1 K.
Thermoelectric voltage $\Delta V$ was measured by a nano-voltmeter (2182A, Keithley).

Figure \ref{fig1}(a) shows the XRD patterns of the single crystalline sample of Nd$_{1-x}$Ce$_{x}$OBiS$_2$ along the $c$-axis at room temperature.
The presence of only 00$l$ diffraction peaks indicates that the $ab$-plane is well grown.
Figure \ref{fig1}(c) shows that the lattice constant $c$ was 1.3661 nm for Nd$_{0.8}$Ce$_{0.2}$OBiS$_2$, and decreased as Ce content \textit{x} increased. 
The result conflicts with lanthanide contraction for a trivalent Ce valence, because the ionic radius of Nd$^{3+}$ is smaller than that of Ce$^{3+}$.
However, the lattice constants of CeOBiS$_2$ are slightly smaller than those of LaOBiS$_2$ and PrOBiS$_2$, indicating that the Ce valence is a mixed state of Ce$^{3+}$/Ce$^{4+}$\cite{CeOBiS2}.
Thus, the shrinking of the lattice constants of the Ce-substituted NdOBiS$_2$ indicates the presence of Ce$^{4+}$.

The XRF analysis helped determine the chemical composition ratios of the single crystals for \textit{x} = 0.2, 0.3, 0.4, and 0.5, which were Nd$_{0.73}$Ce$_{0.24}$BiS$_{1.7}$, Nd$_{0.72}$Ce$_{0.30}$BiS$_{1.9}$, Nd$_{0.59}$Ce$_{0.41}$BiS$_{2.0}$, and Nd$_{0.57}$Ce$_{0.53}$BiS$_{2.1}$, respectively.
The O composition ratio could not be determined.
The values suggest that the chemical composition of the grown crystals is almost consistent with the nominal one, although the ratio of \textit{x} = 0.5 is slightly deviated.
The ratio of the single crystal for \textit{x} = 0.2 seems to indicate a lack of S.
One possible explanation is that the deviation is ascribed to the lumpy shape, which causes a measurement error.
Another possibility is that a lack of S leads to the lumpy shape.

Figures \ref{fig2}(a-b) show the $T$-dependence of the magnetic susceptibility $\chi(T)$ of Nd$_{1-x}$Ce$_{x}$OBiS$_2$ on heating after zero-field cooling (ZFC) and then on cooling in the field (field cooling, FC) for $H$ $||$ $ab$ and $H$ $||$ $c$.
A large diamagnetism was observed in Nd$_{1-x}$Ce$_{x}$OBiS$_{2}$ (\textit{x} = 0.2, 0.3, 0.4, and 0.5).
The significant exceedance of $-$4$\pi\chi$ is due to the diamagnetic factor when $H$ was applied along the $c$-axis.
By applying a magnetic field along the $ab$-plane, the diamagnetic factor is almost negligible.
Figure \ref{fig2}(b) shows that a large diamagnetism was observed in Nd$_{0.8}$Ce$_{0.2}$OBiS$_{2}$ at 4.7 K, suggesting that the superconductivity is of bulk nature, although the shielding effect slightly exceeds $-$1 in 4$\pi\chi$.
It is likely that the roughness of the crystal shape leads to the misalignment of the field applied to the $ab$-plane.

Figures \ref{fig2}(c-d) show that $T_\mathrm{c}$ is defined as the temperature where the magnetization drop starts to appear.
Because $\chi(T)$ for \textit{x} = 0.5 gradually decreases, we consider $T_\mathrm{c}$ as the temperature where FC starts to deviate from the ZFC.
The values obtained for $T_\mathrm{c}$ are 4.75 (\textit{x} = 0.2), 4.55 (\textit{x} = 0.3), 4.42 (\textit{x} = 0.4), and 3.85 K (\textit{x} = 0.5).
$T_\mathrm{c}$ slightly decreases with increasing x.
This behavior corresponds to the F-doped NdOBiS$_2$\cite{5}, which exhibits the highest $T_\mathrm{c}$ (= 5.2 K) at ambient pressure.
In addition, the maximum $T_\mathrm{c}$ of the Ce-doped compounds is slightly lower than that of NdO$_{1 - x}$F$_x$BiS$_{2}$\cite{5}.
Such behavior is also observed in Ce-doped LaOBiSSe and PrOBiS$_{2}$\cite{LaCe,PrCe}.
The origin of the difference in $T_\mathrm{c}$ is not clear.

\begin{figure}
\begin{center}
\includegraphics[width=3.3in]{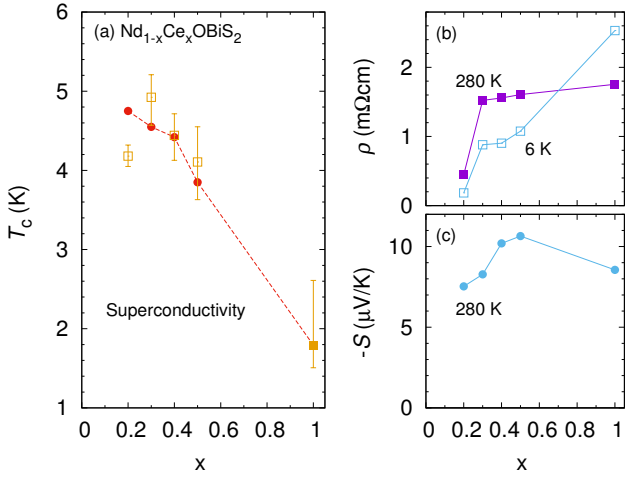}
\caption{(Color Online) 
(a) Superconducting phase diagram of Nd$_{1-x}$Ce$_{x}$OBiS$_2$ determined by the magnetic susceptibility $\chi(T)$, $\bullet$ and the electrical resistivity $\rho(T)$, $\square$.
(b) $\rho(T)$ of Nd$_{1-x}$Ce$_{x}$OBiS$_2$ at $T$ = 5 and 300 K.
The data of CeOBiS$_2$ are extracted from Ref. 27.
The $y$-axis is a logarithmic scale.
}
\label{fig3}
\end{center}
\end{figure}

Figures \ref{fig2}(e-f) show the $T$-dependence of the electrical resistivity $\rho(T)$ of Nd$_{1-x}$Ce$_{x}$OBiS$_2$ at low temperatures.
Superconductivity was confirmed by $\rho(T)$ in all the samples.
$T_\mathrm{c}$ obtained from $\rho(T)$ was determined as the temperature of 50\%$\rho_0$, where $\rho_0$ is the residual resistivity.
The error bar as the superconducting transition width of $T_\mathrm{c}^\mathrm{onset}$ and $T_\mathrm{c}^\mathrm{zero}$ is defined as 98\%$\rho_0$ and 2\%$\rho_0$.
The values obtained for $T_\mathrm{c}$ are 4.18 (\textit{x} = 0.2), 4.92 (\textit{x} = 0.3), 4.47 (\textit{x} = 0.4), and 4.11 K (\textit{x} = 0.5).
Nd$_{0.7}$Ce$_{0.3}$OBiS$_2$ exhibits the highest $T_\mathrm{c}$ among the Ce-doped compounds, where $T_\mathrm{c}^\mathrm{zero}$ is almost consistent with $T_\mathrm{c}$ obtained from $\chi(T)$.
$T_\mathrm{c}$ for \textit{x} = 0.2 obtained from $\rho(T)$ differs from that of $\chi(T)$.
This could be owing to the Joule heat caused when current was applied in an extremely small sample.
A superconducting phase diagram summarizes the above in Fig. \ref{fig3}(a).

Figure \ref{fig2}(f) shows the $T$-dependence of the electrical resistivity $\rho(T)$ of Nd$_{1-x}$Ce$_{x}$OBiS$_2$ down to 3.0 K.
A large hump was observed in $\rho(T)$ of non-doped CeOBiS$_2$ at approximately 130 K.
The result corresponds to the previous report\cite{CeOBiS2}.
Metallic behavior was observed when \textit{x} = 0.2, 0.3, 0.4, and 0.5, indicating that the electron carriers were doped with the mixed-valence state of the Ce ion.
At \textit{x} = 0.5, a small hump was observed at approximately 80 K and was greatly suppressed from the non-doped compound.
Figure \ref{fig3}(b) shows the values of $\rho$(6 K) and $\rho$(280 K) plotted.
Both $\rho$(6 K) and $\rho$(280 K) increase with the Ce content \textit{x}.

To understand the carrier density of the Ce-doped NdOBiS$_2$, we measured the Seebeck coefficient $S$ at 280 K, as shown in Fig. \ref{fig3}(c).
In metals or degenerate semiconductors (single parabolic band and energy-independent scattering approximation)\cite{ST,ST1}, $S/T$ is proportional to 1/$n$, where $n$ is the carrier concentration.
The absolute value of $S$ in \textit{x} = 0.2 is 7.4 $\mu$V/K, which is the smallest value in the Ce-doped NdOBiS$_2$.
With an increase in the Ce content \textit{x}, the absolute value increases, which corresponds to the increase in $\rho$(280 K).
The results show a decrease in the carrier density despite the increase in the amount of Ce substitution.
The behavior can be understood by the changing ratio of Ce$^{3+}$/Ce$^{4+}$ depending on the substitution amount, and the ratio of Ce$^{4+}$ increases as the \textit{x} value decreases.
In contrast, when the Ce concentration is further increased above $x$ = 0.5, the absolute value of $S$ decreases.
The decreasing $S$ conflicts with the $\rho(T)$ behavior in CeOBiS$_2$.
Because the anomaly of $\rho(T)$ is not similar to the metallic or semiconducting behavior, there is a possibility that the assumption of the metals or degenerate semiconductors is inadequate in CeOBiS$_2$.

\begin{figure}
\begin{center}
\includegraphics[width=3.3in]{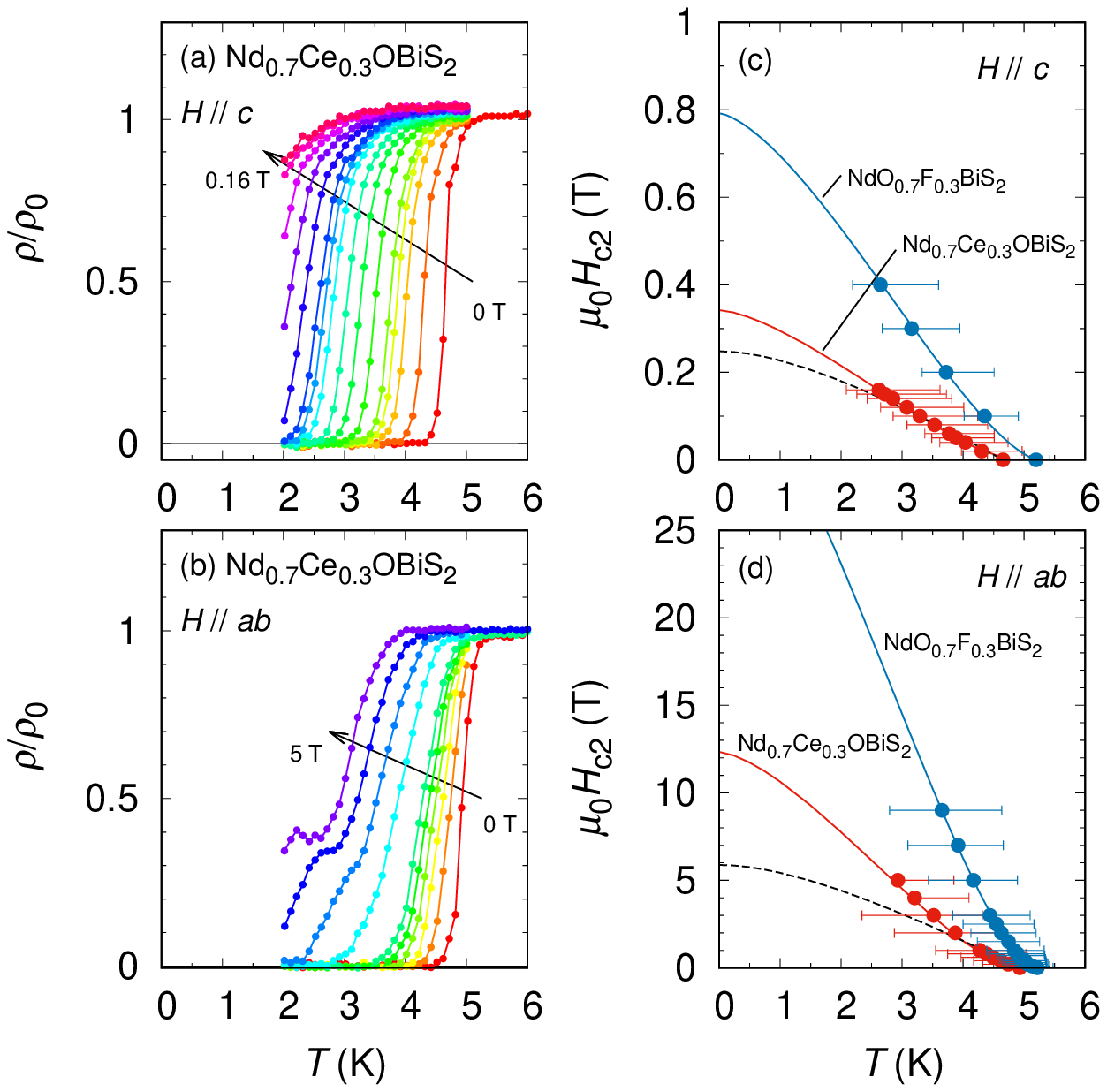}
\caption{(Color Online)
$T$-dependence of the electrical resistivity $\rho(T)$ in several magnetic fields for (a) $H$ $||$ $ab$ and (b) $H$ $||$ $c$.
$T$-dependence of the upper critical field $H_\mathrm{c2}(T)$ of Nd$_{0.7}$Ce$_{0.3}$OBiS$_2$ and NdO$_{0.7}$F$_{0.3}$BiS$_2$ for (c) $H$ $||$ $ab$ and (d) $H$ $||$ $c$.
The data on NdO$_{0.7}$F$_{0.3}$BiS$_2$ were taken from Ref. 30.
The dotted line indicates the Werthamer--Helfand--Hohenberg (WHH)  theory with the dirty limit ($h^\ast$ = 0.69), where $h^\ast$ is $H_\mathrm{c2}$ normalized by the initial slope in the vicinity of $T_\mathrm{c}$.
The solid lines represent the fitting results obtained using $H_\mathrm{c2}(T)$ = $H_\mathrm{c2}$(0)[1 − ($T/T_\mathrm{c}$)$^{3/2}$]$^{3/2}$.
}
\label{fig4}
\end{center}
\end{figure}

Figures \ref{fig4}(a-b) show $\rho(T)$ of Nd$_{0.7}$Ce$_{0.3}$OBiS$_2$ in several magnetic fields applied along with $ab$ and $c$.
The upper critical field $\mu_0H_\mathrm{c2}(T)$ of Nd$_{0.7}$Ce$_{0.3}$OBiS$_2$ is obtained from the $\rho(T)$ measurements in several magnetic fields, as shown in Figs. \ref{fig4}(c-d).
The superconductivity is significantly suppressed with increasing magnetic fields in both directions.
$\rho(T)$ above 3 T shows a small hump at a low temperature point, which is not reproducible.
It is likely that the behaviors are not intrinsic.
As shown in Figs. \ref{fig4}(c-d), the dotted lines show the Werthamer--Helfand--Hohenberg (WHH) theory prediction\cite{hc2b,hc2a}.
In both curves, the $H_\mathrm{c2}(T)$ curves deviate from the WHH prediction and exhibit a positive curvature at around $T_\mathrm{c}$.
This behavior is mainly attributed to spin-orbit coupling\cite{hc2a}, multi-gap superconductivity\cite{multi1,multi}, and strong electron--phonon coupling\cite{hc2,strong}.
However, the origin of the positive curvature in the BiS$_2$-based superconductors still remains to be clarified.
A reasonable fit to the data for Nd$_{0.7}$Ce$_{0.3}$OBiS$_2$ can be obtained using the expression $H_\mathrm{c2}(T)$ = $H_\mathrm{c2}$(0)[1 − ($T/T_\mathrm{c}$)$^{3/2}$]$^{3/2}$\cite{hc2fit}.
This model was also used to fit $H_\mathrm{c2}(T)$ for BiS$_2$-based superconductors\cite{spe,Bi4O4S3}, (Zr, Hf)IrSi\cite{TrIrSi}, Nb$_{0.18}$Re$_{0.82}$\cite{NbRe}, WRe$_3$\cite{WRe3}, PbTaSe$_2$\cite{PbTaSe2}, and high-entropy alloys\cite{entro}.
$\mu_0H_\mathrm{c2}(T)$ was determined as $\sim$12 T for the $H$ $||$ $ab$ and $\sim$0.34 T for the $H$ $||$ $c$-axis.
In this article, we redetermined $H_\mathrm{c2}(T)$ of NdO$_{0.7}$F$_{0.3}$BiS$_2$ from the same definition of $T_\mathrm{c}$, as shown in Figs. \ref{fig4}(c-d). 
The data were taken from the previous report\cite{NdOBiS2}.
$\mu_0H_\mathrm{c2}(0)$ was determined as $\sim$35 T for the $H$ $||$ $ab$ and $\sim$0.78 T for the $H$ $||$ $c$-axis.
These values are clearly larger than those of the Ce-doped NdOBiS$_2$.
The results indicate that the $\mu_0H_\mathrm{c2}(0)$ for both directions reduces in superconductivity induced by the Ce-doping.
Because measurements were performed above 2.0 K and above 9 T, lower temperature and higher field measurements were needed to determine the accurate $H_\mathrm{c2}$ values of the F- and Ce-doped NdOBiS$_2$.

The anisotropic parameter $\Gamma$ was determined using the anisotropic Ginzburg--Landau (GL) formula,
\begin{equation}
\label{ }
\Gamma = \frac{m_c}{m_{ab}} = \frac{H^{|| ab}_\mathrm{c2}}{H^{|| c}_\mathrm{c2}} = \frac{\xi_{ab}}{\xi_c},
\end{equation}
where $m_{ab}$, $m_{c}$ are the effective masses, and $\xi_{ab}$, $\xi_{c}$ are the GL-coherence lengths.
$\xi_{ab}$ and $\xi_c$ were evaluated using $H^{|| ab}_\mathrm{c2}$ = $\Phi_0$/(2$\pi\xi_{ab}\xi_c$) and $H^{|| c}_\mathrm{c2}$ = $\Phi_0$/(2$\pi\xi_{ab}^2$), where $\Phi_0$ is the quantum flux.
$\xi_{ab}$ and $\xi_c$ were obtained as $\sim$86 and 0.31 nm for Nd$_{0.7}$Ce$_{0.3}$OBiS$_2$, respectively.
The anisotropic parameter $\Gamma$ was estimated to be $\sim$35.
The result suggests that the superconductivity possesses high anisotropy.
The $\Gamma$ of NdO$_{0.7}$F$_{0.3}$BiS$_2$ was recalculated to be $\sim$45.
Because the difference in $\Gamma$ is approximately 10, there is no considerable difference between F-doped and Ce-doped NdOBiS$_2$.

In summary, we succeeded in growing a single crystal of Nd$_{1-x}$Ce$_{x}$OBiS$_{2}$ (\textit{x} = 0.2, 0.3, 0.4, and 0.5).
Plate-like single crystals with dimensions of 0.8$\times$0.8$\times$0.2 mm$^3$ were obtained.
We discovered bulk superconductivity at $T_\mathrm{c}$ = 4.7 K in a compound with \textit{x} = 0.2.
The lattice constant of the $c$-axis decreased with an increase in Ce content \textit{x}.
The shrinking of the $c$-axis indicated the mixed valency of Ce$^{3+}$/Ce$^{4+}$.
It is likely that the superconductivity of Nd$_{1-x}$Ce$_{x}$OBiS$_{2}$ was induced by the electron doping of the mixed valence of Ce$^{3+}$/Ce$^{4+}$.
The superconducting transition temperature of Ce-substituted Nd$_{1-x}$Ce$_{x}$OBiS$_{2}$ was slightly lower than that of F-substituted NdO$_{1 - x}$F$_{x}$BiS$_{2}$.
This tendency is also observed in Ce-substituted La$_{1 - x}$Ce$_{x}$OBiSSe and Ce$_{1 - x}$Pr$_{x}$OBiS$_{2}$.
$\rho(T)$ of Nd$_{1-x}$Ce$_{x}$OBiS$_{2}$ showed a metallic behavior.
In a compound with \textit{x} = 0.5, a broad hump was observed in $\rho(T)$ at approximately 80 K, similar to that found in CeOBiS$_2$.
The electrical resistivity of $\rho$(6 K) and $\rho$(280 K) increased with increasing \textit{x}.
The tendency of $S$ can be understood by the increase of the Ce$^{4+}$ ratio with decreasing Ce content below \textit{x} = 0.5.
The upper critical field $\mu_0H_\mathrm{c2}(T)$ of Nd$_{0.7}$Ce$_{0.3}$OBiS$_{2}$ was determined to be $\sim$12 T for $H$ $||$ $ab$ and $\sim$0.34 T for $H$ $||$ $c$-axis.
The anisotropic parameter $\Gamma$ was estimated to be $\sim$35, which was comparable to that of F-doped NdOBiS$_2$.
The $\mu_0H_\mathrm{c2}$(0) of the Ce-doped NdOBiS$_2$ was approximately two times smaller than that of the F-doped one, although the difference of $\Gamma$ was approximately 10.
The results imply that the $\mu_0H_\mathrm{c2}(0)$ for both directions reduces in superconductivity induced by Ce-doping.
Our results for Ce-doped NdOBiS$_2$ make it possible to help understand superconductivity in BiS$_2$-based superconductors.

\section*{Acknowledgment}
This work was partially supported by the Kurata Grants from the Hitachi Global Foundation, Izumi Science and Technology Foundation, and Casio Science Promotion Foundation.
The authors would like to thank Prof. Ryuji Tamura for his help in measuring the XRF data.
DC-magnetic susceptibility performed by MPMS and electrical resistivity in magnetic fields by PPMS were carried out in collaboration with the Institute for Solid State Physics, University of Tokyo.


\begin{thebibliography}{9}
\bibitem{1} Y. Mizuguchi, H. Fujishita, Y. Gotoh, K. Suzuki, H. Usui, K. Kuroki, S. Demura, Y. Takano, H. Izawa, and O. Miura, Phys. Rev. B, \textbf{86}, 220510(R) (2012).
\bibitem{2} Y. Mizuguchi, S. Demura, K. Deguchi, Y. Takano, H. Fujishita, Y. Gotoh, H. Izawa, and O. Miura, J. Phys. Soc. Jpn., \textbf{81}, 114725 (2012).
\bibitem{3} J. Xing, S. Li, X. Ding, H. Yang, and H.-H. Wen, Phys. Rev. B, \textbf{86}, 214518 (2012).
\bibitem{4} D. Yazici, K. Huang, B. D. White, A. H. Chang, A. J. Friedman, and M. B. Maple, Philos. Mag., \textbf{93}, 673 (2013).
\bibitem{5} S. Demura, Y. Mizuguchi, K. Deguchi, H. Okazaki, H. Hara, T. Watanabe, S. J. Denholme, M. Fujioka, T. Ozaki, H. Fujihisa, J. Phys. Soc. Jpn., \textbf{82}, 033708 (2013).
\bibitem{6} R. Jha, A. Kumar, S. Kumar Singh, and V. P. S. Awana, J. Supercond. Nov. Magn., \textbf{26}, 499 (2013).
\bibitem{7} X. Lin, X. Ni, B. Chen, X. Xu, X. Yang, J. Dai, Y. Li, X. Yang, Y. Luo, Q. Tao, G. Cao, and Z. Xu, Phys. Rev. B \textbf{87}, 020504(R) (2013)
\bibitem{8} R. Jha, B. Tiwari, and V. P. S. Awana, J. Applied Physics \textbf{117}, 013901 (2015).
\bibitem{9} R. Jha, B. Tiwari, and V. P. S. Awana, J. Phys. Soc. Jpn., \textbf{83}, 063707 (2014).
\bibitem{theory1} H. Usui, K. Suzuki, and K. Kuroki, Phys. Rev. B \textbf{86}, 220501 (2012).
\bibitem{theory2} H. Usui and K. Kuroki, Nov. Supercond. Mater., \textbf{1}, 50 (2015).
\bibitem{a0} Y. Mizuguchi, A. Miura, J. Kajitani, T. Hiroi, O. Miura, K. Tadanaga, N. Kumada, E. Magome, C. Moriyoshi, and Y. Kuroiwa, Sci. Rep. \textbf{5}, 14968 (2015).
\bibitem{a1} T. Hiroi, J. Kajitani, A. Omachi, O. Miura, and Y. Mizuguchi, J. Phys. Soc. Jpn. \textbf{84}, 024723 (2015).
\bibitem{a2} G. Jinno, R. Jha, A. Yamada, R. Higashinaka, T. D. Matsuda, Y. Aoki, M. Nagao, O. Miura, and Y. Mizuguchi, J. Phys. Soc. Jpn. \textbf{85}, 124708 (2016).
\bibitem{a3} Y. Goto, R. Sogabe, and Y. Mizuguchi, J. Phys. Soc. Jpn. \textbf{86}, 104712 (2017).
\bibitem{b0} E. Paris, B. Joseph, A. Iadecola, T. Sugimoto, L. Olivi, S. Demura, Y. Mizuguchi, Y. Takano, T. Mizokawa, and N. L. Saini, J. Phys.: Condens. Matter, \textbf{26} 435701 (2014).
\bibitem{b1} Y. Mizuguchi, E. Paris, T. Sugimoto, A. Iadecola, J. Kajitani, O. Miura, T. Mizokawa and N. L. Saini, Phys. Chem. Chem. Phys. \textbf{17}, 22090 (2015).
\bibitem{b2} A. Athauda, J. Yang, S. Lee, Y. Mizuguchi, K. Deguchi, Y. Takano, O. Miura, and D. Louca, Phys. Rev. B \textbf{91}, 144112 (2014).
\bibitem{b3} K. Nagasaka, A. Nishida, R. Jha, J. Kajitani, O. Miura, R. Higashinaka, T. D. Matsuda, Y. Aoki, A. Miura, C. Moriyoshi, Y. Kuroiwa, H. Usui, K. Kuroki, and Y. Mizuguchi, J. Phys. Soc. Jpn. \textbf{86}, 074701 (2017).
\bibitem{spe} N. Kase, Y. Terui, T. Nakano, and N. Takeda, Phys. Rev. B  \textbf{96}, 214506 (2017).
\bibitem{spe1} N. Kase, J. Phys. Soc. Jpn., \textbf{88}, 041007 (2019).
\bibitem{iso} K. Hoshi, Y. Goto, and Y. Mizuguchi, Phys. Rev. B \textbf{97}, 094509 (2018).
\bibitem{thermal1} L. Jiao, Z. Weng, J. Liu, J. Zhang, G. Pang, C. Guo, F. Gao, X. Zhu, H.-H. Wen, and H. Q. Yuan, J. Phys.: Condens. Matter \textbf{27}, 225701 (2015).
\bibitem{thermal2} Shruti, S. P. Srivastava, and S. Patnaik, J. Phys.: Condens. Matter \textbf{25}, 312202 (2013).
\bibitem{thermal3} T. Yamashita, Y. Tokiwa, D. Terazawa, M. Watauchi, I. Tanaka, T. Terashima, and Y. Matsuda, J. Phys. Soc. Jpn. \textbf{85}, 073707 (2016).
\bibitem{ARPES} Y. Ota, K. Okazaki, H. Q. Yamamoto, T. Yamamoto, S. Watanabe, C. Chen, M. Nagao, S. Watauchi, I. Tanaka, Y. Takano, and S. Shin, Phys. Rev. Lett. \textbf{118}, 167002 (2017).
\bibitem{CeOBiS2} M. Tanaka, M. Nagao, R. Matsumoto, N. Kataoka, I. Ueta, H. Tanaka, S. Watauchi, I. Tanaka, and Y. Takano, J. Alloys Compd., \textbf{722}, (2017) 467. 
\bibitem{LaCe} R. Sogabe, Y. Goto, A. Nishida, T. Katase, and Y. Mizuguchi, EPL, \textbf{122}, (2018) 17004.
\bibitem{PrCe} A. Miura, M. Nagao, Y. Goto, Y. Mizuguchi, T. D. Matsuda, Y. Aoki, C. Moriyoshi, Y. Kuroiwa, Y. Takano, S. Watauchi, I. Tanaka, N. C. Rosero-Navarro, and K. Tadanaga, Inorg. Chem. \textbf{57}, 5364-5370 (2018).
\bibitem{NdOBiS2} M. Nagao, S. Demura, K. Deguchi, A. Miura, S. Watauchi, T. Takei, Y. Takano, N. Kumada, and I. Tanaka, J. Phys. Soc. Jpn., \textbf{82}, 113701 (2013). 
\bibitem{supp} (Supplemental material) Detailed results of the electrical resistivity are provided online.
\bibitem{ST} M. Cutler, J. F. Leavy, and R. L. Fitzatrick, Phys. Rev., \textbf{133} (1964) A1143.
\bibitem{ST1} G. J. Snyder and E. S. Toberer, Nat. Mater., \textbf{7}, 105-114 (2008).
\bibitem{hc2b} K. Maki, Phys. Rev., \textbf{148}, (1966) 362.
\bibitem{hc2a} E. Helfand and N. R. Werthamer, Phys. Rev., \textbf{147}, (1966) 288.
\bibitem{multi1} Y. Nakajima, H. Hidaka, T. Nakagawa, T. Tamegai, T. Nishizaki, T. Sasaki, N. Kobayashi, Phys. Rev. B \textbf{85}, 174524 (2012).
\bibitem{multi} Y. C. Chan, K. Y. Yip, Y. W. Cheung, Y. T. Chan, Q. Niu, J. Kajitani, R. Higashinaka, T. D. Matsuda, Y. Yanase, Y. Aoki, K. T. Lai, and Swee K. Goh, Phys. Rev. B \textbf{97}, 104509 (2018).
\bibitem{hc2} L. N. Bulaevskii, O. V. Dolgov, and M. O. Ptitsyn, Phys. Rev. B, \textbf{38}, 11290 (1988).
\bibitem{strong} F. Marsiglio and J. P. Carbotte, Phys. Rev. B \textbf{41}, 8765 (1990).
\bibitem{hc2fit} R. Micnas, J. Ranninger, and S. Robaszkiewicz, Rev. Mod. Phys., \textbf{62}, 113 (1990).
\bibitem{Bi4O4S3} P. K. Biswas, A. Amato, C. Baines, R. Khasanov, H. Luetkens, H. Lei, C. Petrovic, and E. Morenzoni, Phys. Rev. B \textbf{88}, 224515 (2013).
\bibitem{TrIrSi} N. Kase, H. Suzuki, T. Nakano, and N. Takeda, Supercond. Sci. Technol. \textbf{29}, 035011 (2016).
\bibitem{NbRe} A. B. Karki, Y. M. Xiong, N. Haldolaarachchige, S. Stadler, I. Vekhter, P. W. Adams, D. P. Young, W. A. Phelan, and J. Y. Chan, Phys. Rev. B \textbf{83}, 144525 (2011).
\bibitem{WRe3} P. K. Biswas, M. R. Lees, A. D. Hillier, R. I. Smith, W. G. Marshall, and D. McK. Paul, Phys. Rev. B \textbf{84}, 184529 (2011).
\bibitem{PbTaSe2} M. N. Ali, Q. D. Gibson, T. Klimczuk, and R. J. Cava, Phys. Rev. B, \textbf{89}, 020505 (2014).
\bibitem{entro} K. Stolze, J. Tao, F. O. von Rohr, T. Kong, and R. J. Cava, Chem. Mater. \textbf{30}, 906-914 (2018).

\end{thebibliography}
\end{document}